\documentstyle[12pt,aas2pp4]{article}

\begin{document}

\title{Anticorrelated Hard/Soft X-ray Emission from the X-Ray 
Burster 4U~0614+091}

\authoremail{eric@orphee.phys.columbia.edu}
\author{E. Ford\altaffilmark{1}, P. Kaaret\altaffilmark{1},
   M. Tavani\altaffilmark{1},
   B.A. Harmon\altaffilmark{2},S.N. Zhang\altaffilmark{3},
   D. Barret\altaffilmark{4}, J. Grindlay\altaffilmark{4},
   P. Bloser\altaffilmark{4}, R. A. Remillard\altaffilmark{5}}
\altaffiltext{1}{Columbia University, Department of Physics and
 Columbia Astrophysics Lab, 538 W. 120th Street, New York, NY 10027}
\altaffiltext{2}{NASA/Marshall Space Flight Center,
 ES 84, Huntsville, AL 35812}
\altaffiltext{3}{University Space Research Association/ MSFC,
 ES 84, Huntsville, AL 35812}
\altaffiltext{4}{Harvard Smithsonian Center for Astrophysics,
 60 Garden Street, Cambridge, MA 02138}
\altaffiltext{5}{MIT, Center for Space Science, Room 37-595,
 Cambridge, MA 02139}

\begin{abstract}

We have detected transient X-ray activity from the X-ray burster 4U~0614+091 
simultaneously with BATSE/CGRO (20--100 keV) and ASM/RXTE (1--12 keV). The 
peak fluxes reach approximately 40 mCrab in both instruments over a period
of about 20 days. The variable emission shows a clear anticorrelation of 
the hard X-ray flux with the soft X-ray count rate. The observed 
anticorrelation is another clear counterexample to the notion that only 
black hole binaries exhibit such correlations. The individual spectra during 
this period can be fit by power laws with photon indices $2.2\pm0.3$ (ASM) 
and $2.7\pm0.4$ (BATSE), while the combined spectra can be described by a 
single power law with index $2.09\pm0.08$. BATSE and the ASM/RXTE are a good 
combination for monitoring X-ray sources over a wide energy band.

\end{abstract}


\section{Introduction}

Many X-ray emission properties once thought to uniquely characterize black 
hole candidates have now been observed in systems containing weakly magnetized
neutron stars (e.g. Barret and Vedrenne 1994; van der Klis 1994; Inoue 1990). 
A classic observational feature of black hole candidates (BHCs) has been the 
presence of hard power law tails in the energy spectra extending above 
100 keV. Typically the hard X-ray emission is anti-correlated with soft X-ray 
luminosity (Tanaka 1989). Measurements below 20 keV show that X-ray bursters,
representing weakly magnetized neutron stars (NSs), also have power law tails.
As in the BHCs, there is a hardness/flux anticorrelation, clearly observed in 
a number of bursters: e.g. 4U~1636-536 (Breedon et al. 1986), 
4U~1735-444 (Smale et al. 1986), 4U~1705-44 
(Langmeier et al. 1987), 4U~1608-522 (Mitsuda et al. 1989) and 4U~0614+091 
(Barret and Grindlay 1995).

Until recently, most observations of accreting NSs have been limited to less 
than 20 keV. A notable exception was the detection of a 150 day outburst from
the burster Cen~X-4 with SIGNE 2 MP/PROGNOZ 7 (Bouchacourt et al. 1984) in the
3--163 keV band. In this observation the hard X-ray flux was high both at the 
beginning and end of the outburst when the soft X-ray flux was low. With 
observations by SIGMA, the situation has changed and many more X-ray bursters
are now detected above 20 keV (Barret and Vedrenne 1994). The hard X-ray 
emission is typically variable on the time scale of days and seems to be 
detected preferentially from the low luminosity bursters. Additional bursters 
are being detected in hard X-rays with BATSE monitoring (see Barret et al. 
1996). Numerous X-ray bursters are now firmly established as sources of hard 
X-ray emission.

4U~0614+091 is one of the low luminosity bursters. It was previously 
identified as an X-ray burster (Swank et al. 1978; Brandt et al. 1992; 
Brandt 1994) and as a probable atoll source (Singh and Apparao 1995). BATSE 
monitoring has revealed numerous episodes of transient emission above 20 keV
from 4U~0614+091 (Ford et al. 1996a).

Here we describe the simultaneous observations of 4U~0614+091 using the
Burst and Transient Source Experiment (BATSE) aboard the Compton Gamma-Ray
Observatory (CGRO) and the All Sky Monitor (ASM) onboard the Bruno B. Rossi
X-ray Timing Explorer (RXTE). In April 1996, with near real-time BATSE 
monitoring, we discovered an episode of hard X-ray activity from 4U~0614+091 
which was between 4 and 15 days long. This event provided a trigger for RXTE 
pointings on TJD 10195 and 10197, the results of which will be presented 
elsewhere (Ford et al. 1996b). The correlated ASM and BATSE observations 
presented here cover this April 1996 event. In Section 2 we describe the 
observations by both instruments. In Section 3 we discuss the implications of 
these new results.

\section{Observations}

The analysis with BATSE uses the technique of Earth occultation, described 
for example in Harmon et al. (1992) and Zhang et al. (1993). This method 
offers a 20--150 keV $3\sigma$ sensitivity for two week integrations of 
approximately $1.7\times10^{-2}$ ph cm$^{-2}$~s$^{-1}$ (50 mCrab). With 
favorable detector geometries (when a source is near a detector normal) the 
sensitivity can improve to approximately 
$1.5\times10^{-2}$ ph cm$^{-2}$~s$^{-1}$ in only 2 days.

The ASM has been fully operational since early February 1996 ($\sim$TJD 10132), 
following the launch of RXTE in December 1995. The ASM consists of three 
Scanning Shadow Cameras (SSCs) which scan the sky about 7 times each day.
The ASM detectors are position-sensitive proportional counters
placed below coded masks.  X-ray intensity measurements are derived from the
deconvolution of overlapping mask shadows from each X-ray source in the
the field of view.  The position histograms for each camera are 
separately recorded in 3 energy bands (1.3-3.0 keV, 3.0-4.8 keV, and 
4.8-12.2 keV).  The data are processed through standard filters such as a 
$75^\circ$ cut on the Earth angle. In addition, for this analysis we consider 
only on-source dwells in which 4U~0614+091 is the only source in the field of 
view. The fine details of the instrument calibration are still being refined, 
and the present analysis is subject to a systematic error of about 3\% for 
uncrowded source regions as inferred from the rms deviations observed in the 
Crab Nebula. Further details of the instrument and the data analysis methods 
are provided by Levine et al. (1996).

\begin{figure*}
\figurenum{1}
\epsscale{2.0}
\plotone{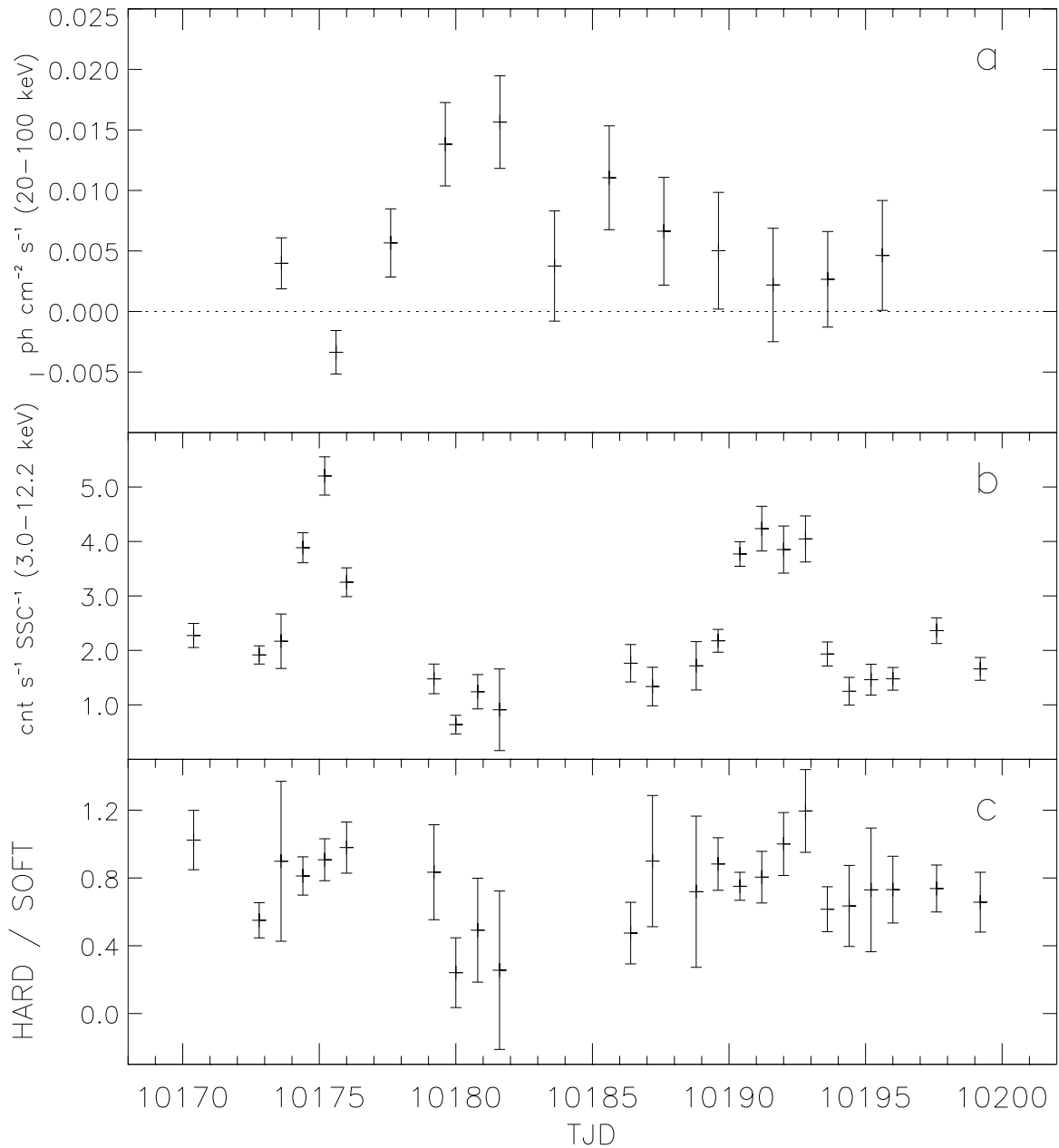}
\caption{Combined BATSE ({\it a}) and ASM/RXTE ({\it b}) light curves. The 
BATSE and ASM data are binned in 2 and 0.8 day bins respectively. The 
BATSE rate to flux conversion assumes a power law spectrum with index 2.8. 
The [4.8--12.0 keV]/[1.3-3.0 keV] hardness ratio from the ASM data is shown in 
({\it c}). Truncated Julian Day (TJD) 10175 corresponds to 2 April 1996.}
\label{fig:lc}
\end{figure*}

\placefigure{fig:lc}

The 4U~0614+091 light curves are shown in Figure~\ref{fig:lc}. Integrating the 
BATSE fluxes (Figure~\ref{fig:lc}a) over TJD 10178--10196, the source is 
detected at a significance of $5.7\sigma$. To confirm that the BATSE light 
curve is not contaminated by interfering sources, we have produced maps of the 
region around 4U~0614+091 during the time of the activity. Maps are generated 
with the occultation technique by producing rate histories over a grid of 
points (see Ford et al. 1996a). The maps show an obvious enhancement centered 
on the position of 4U~0614+091, clearly separated from the Crab which lies 
approximately $17^\circ$ away. At periodic intervals, the limbs passing 
through 4U~0614+091 also sweep over the Crab. This occurs for about 25 days 
prior to TJD 10174 and after TJD 10197. At these times there is no information
on the hard X-ray flux.

Figure~\ref{fig:lc}b shows the ASM light curve (3.0--12.2 keV) binned into 0.8
day averages contemporaneous with the BATSE activity. Two separate outburst 
peaks are clearly visible at TJD 10175 and 10191. The gap in the ASM coverage 
near TJD 10185 was caused by scheduling problems in the RXTE operations. 

\begin{figure*}
\figurenum{2}
\epsscale{2.0}
\plotone{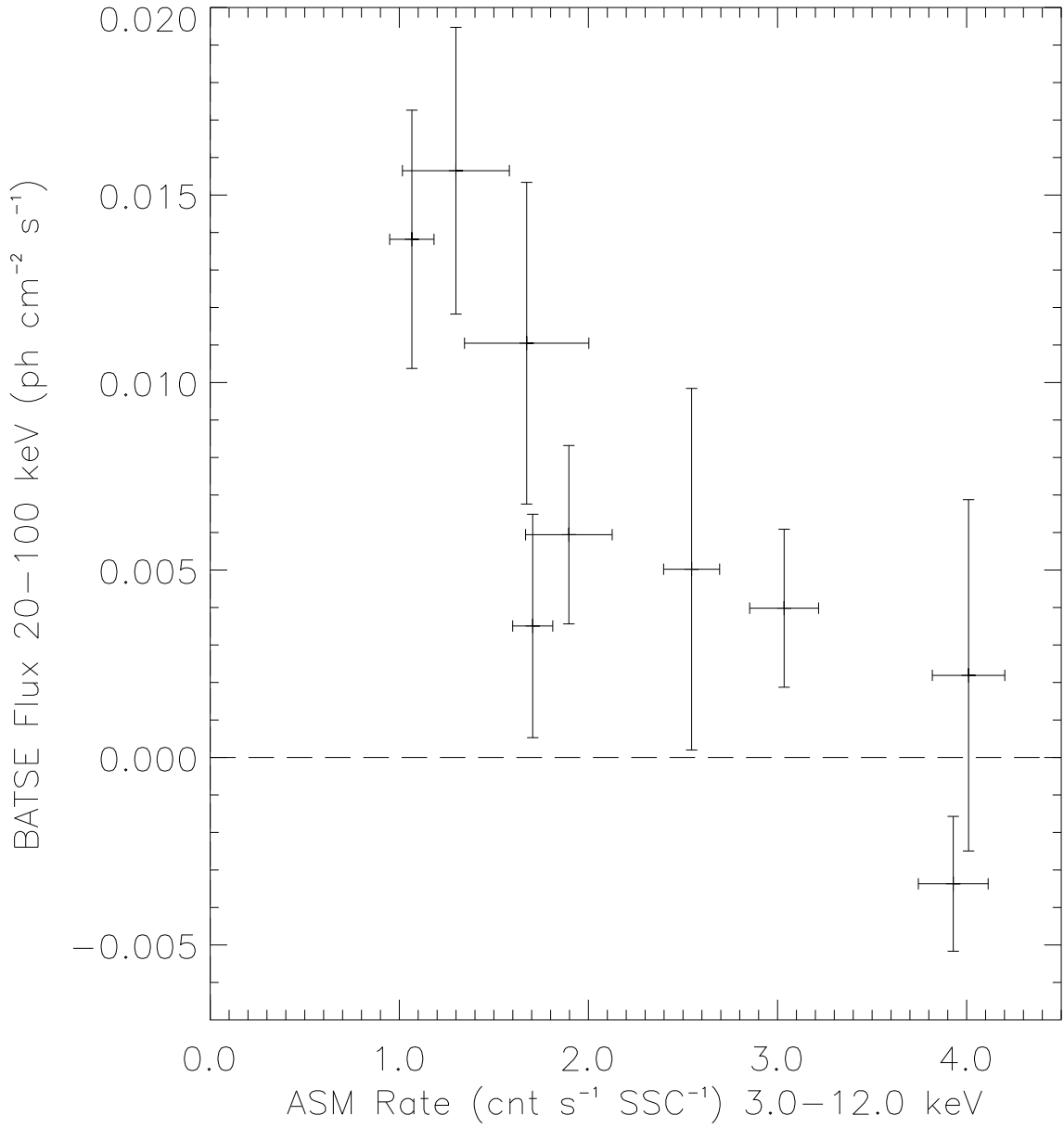}
\caption{BATSE 20--100 keV flux vs ASM 3.0--12.0 keV count rate. 
Data are binned in matching 2 day intervals.}
\label{fig:corel}
\end{figure*}

\placefigure{fig:corel}

A striking feature of Figure~\ref{fig:lc} is the anticorrelation of
the BATSE flux with the ASM count rate. This anticorrelation is exhibited 
more clearly in Figure~\ref{fig:corel}, which uses a common time binning 
of 2.0 days for the hard and soft fluxes. The anticorrelation is robust to
changes in the energy bands.

We can define hardness ratios for the ASM fluxes, using the rates
in the 1.3--3.0, 3.0--4.8 and 4.8--12.0 keV bands ($R1$, $R2$, 
and $R3$ respectively). The $R3/R1$ hardness ratio is shown in 
Figure~\ref{fig:lc}c. $R3/R1$, and also the other hardness ratios, do not show 
a significant correlation with the source intensity. Color-color diagrams of 
the ASM data indicate that we can not rule out a small contribution from 
a 1.0--1.5 keV blackbody as in Barret and Grindlay (1995).

We have produced spectra from the ASM and BATSE data in the window TJD
10178--10190. During this entire interval the source is in a `hard' state;
the ASM count rate is low while the BATSE flux reaches a peak. To generate the 
ASM spectrum, count rates were converted to fluxes by normalizing to the Crab 
rates as observed by the ASM over the same period. The flux errors are 
calculated from the observed (gaussian) distribution of the Crab count rate. 
For a conservative error estimate, we use the 3$\sigma$ value of the Crab
rate distribution. The resulting flux errors for 4U~0614+091 are approximately 
10\% in the log-log space used for spectral fitting. A power law fit to the 
three ASM bins yields a spectral index of $2.23\pm0.33$ ($\chi^{2}_{\nu}=0.5$).
The low $\chi^{2}$ results from the conservative flux error estimate.
The averaged ASM spectrum corresponds to a 1--20 keV flux of 
$1\times10^{-9}$ erg~cm$^{-2}$~s$^{-1}$ with an implied luminosity of 
$1\times10^{36}$ erg~s$^{-1}$ at 3 kpc.

The BATSE data in the same time window can be fit by a power law with photon 
index $2.68\pm0.35$ ($\chi^2_\nu=1.1$). This fit is consistent with the 
spectra from an archival BATSE analysis of 4U~0614+091 during hard X-ray 
emission episodes (Ford et al. 1996a). The BATSE flux is $7\times10^{-10}$ 
erg~cm$^{-2}$~s$^{-1}$ (20--100keV) corresponding to a luminosity of
$8\times10^{35}$ erg~s$^{-1}$ at 3 kpc. 

\begin{figure*}
\figurenum{3}
\epsscale{2.0}
\plotone{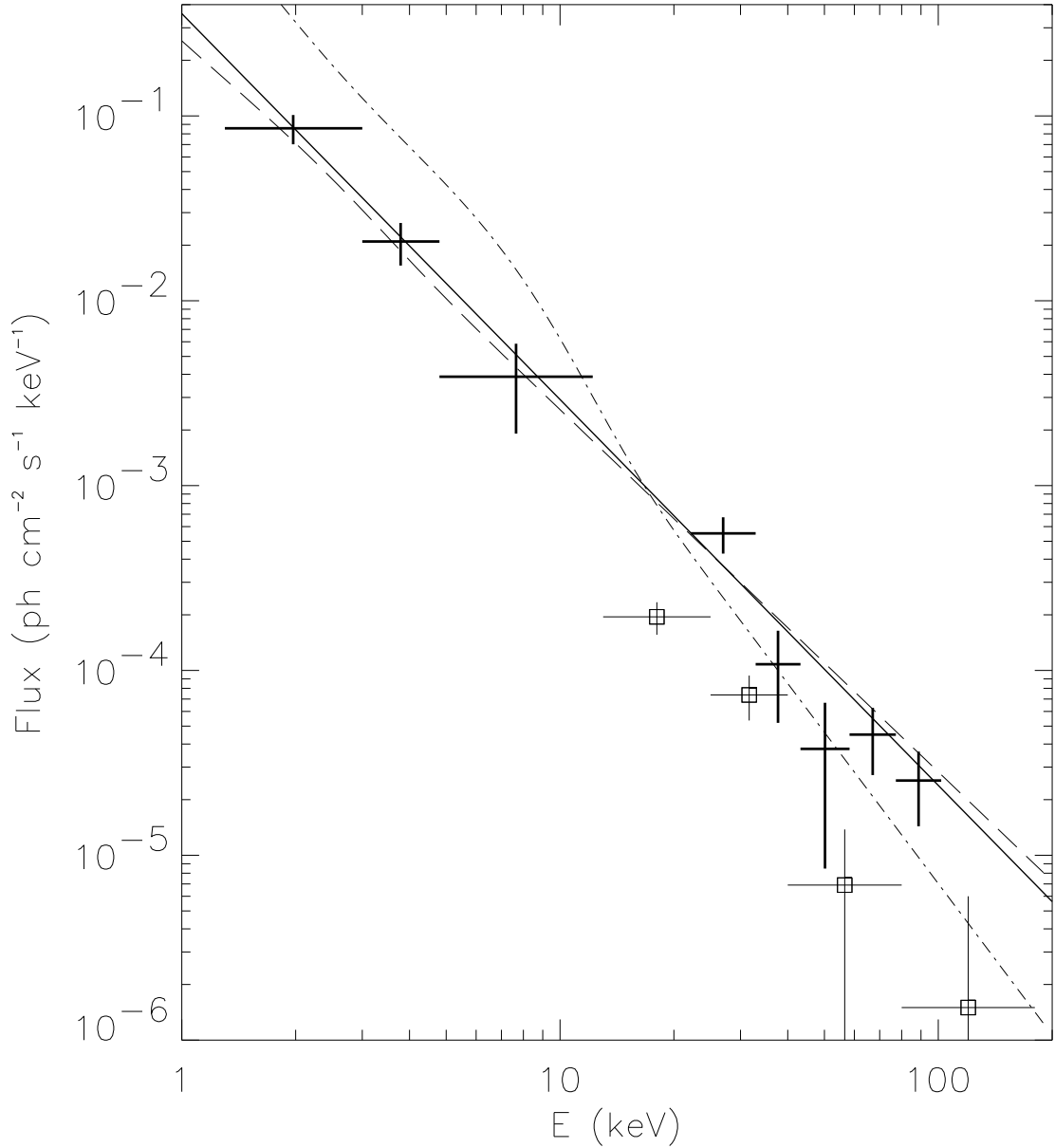}
\caption{Combined BATSE and ASM spectrum over the period TJD 10178--10190.
The solid line is the single power law fit ($\alpha=2.09\pm0.08$) to the
ASM and BATSE fluxes. The dashed (dot--dash) lines show the low (high)
states as identified by EXOSAT. These are extrapolated above the EXOSAT 
limit of 15 keV. Squares are measurements by HEAO-1 A4 from Levine et al. 
(1984).}
\label{fig:spectrum}
\end{figure*}

\placefigure{fig:spectrum}

Figure~\ref{fig:spectrum} shows the combined BATSE and ASM spectra in the
TJD 10178--10190 window. A single power law with index $2.09\pm0.08$
can describe the combined data ($\chi^2_\nu=1.1$). The index is harder than 
either of the two independent fits. We note that this spectral index is
consistent with the the range 1.86--2.24 for the low state as observed
with EXOSAT (Barret and Grindlay 1995).
Although, it is difficult to test more complicated models, 
we attempt to fit a power law with a cutoff, which is an approximation to the 
Comptonization model in the optically thin regime. Fitting results are
summarized in Table~\ref{table:spec_fits}.

\placetable{table:spec_fits}

To test the validity of combining ASM and BATSE spectra we have generated 
spectra for the Crab over the same time interval. The spectra match to
within $\sim20$\%, the BATSE flux being higher than the ASM flux. 
The difference has a low ($\sim1\sigma$) significance. With our conservative 
error estimates, the ASM fluxes are in agreement with the extrapolated BATSE 
fit with a $\chi^2$ of 0.93.

\section{Discussion}

Taken independently, our results in the soft and hard X-ray bands are 
consistent with previous observations. From 1--20 keV, 4U~0614+091 has been 
observed before at similar luminosities; e.g. $5\times10^{36}$~erg~s$^{-1}$
at 3 kpc with EXOSAT (Barret and Grindlay 1995). In the hard X-ray band, 
numerous episodes of emission, similar to this one, have been identified in 
BATSE monitoring (Ford et al. 1996a). Also similar to what we find 
here, variability has been recorded by factors of approximately 5 on time 
scales of days. The 20--100 keV flux extrapolated from the EXOSAT spectrum in 
a high state falls a factor of 4 below the peak of the present hard X-ray 
outburst, and would not be detectable by BATSE. The EXOSAT low state is 
closely matches the BATSE and ASM detections during the hard X-ray episode
(Figure~\ref{fig:spectrum}). 

This observation of 4U~0614+091 is unique for its simultaneous broad-band
coverage. The most notable feature is the clear soft/hard X-ray anticorrelation 
(Figure~\ref{fig:corel}), which confirms a qualitative trend of anticorrelation 
noted earlier at lower energies in 4U~0614+091 and other bursters. The observed 
soft/hard anticorrelation indicates that either the soft or the hard 
X-ray emission is not a direct measure of the mass accretion rate. Analyzing
short segments of ASM and BASTE data, we have found that the total energy 
flux in the 1--100 keV band tracks the 1--12 keV flux, while the hard 
20--100 keV energy flux remains below $1\times10^{-9}$ ergs~cm$^{-2}$~s$^{-1}$
with no observable correlation. This indicates that it is in fact the soft 
X-ray emission which tracks the mass accretion, as is usually assumed. The 
hard X-ray flux then is actually anticorrelated with mass accretion rate, and 
is most likely a manifestation of a special property of the accretion disk. 
We note that in black hole candidates, for example Cyg~X-1 (Crary et al. 1996), 
the hard X-ray emission is also believed to be anticorrelated with the mass 
transfer rates.

Both thermal (Comptonization) and non-thermal models of emission can be
considered in explaining the hard X-ray emission properties of 4U~0614+091.
A thermal model (e.g. Sunyaev and Titarchuk 1980) could account for the
hard/soft anticorrelation in the following way: when the soft X-ray flux
increases, the Comptonizing plasma cools, resulting in a decreased temperature
and hard X-ray flux. 
Non-thermal models have also predicted hard X-ray emission from neutron stars
(Kluzniak et al. 1988) and may explain our results. In non-thermal models,
the soft/hard X-ray anticorrelation is the result of the suppression of
particle acceleration by inverse Compton cooling caused by an enhanced
soft X-ray background. Only for intermediate/low values of $L_s$ can the
hard X-ray emission manifest itself as a consequence of a non-thermal
particle energy distribution function of energized particles
in the disk (Tavani and Liang 1996).

These 4U~0614+091 results are a complement to recent observations of the X-ray 
burster 4U~1608-522 (Zhang et al. 1996). 4U~1608-522 showed a bright
($\sim100$ mCrab) outburst in the BATSE light curve lasting approximately 200 
days. During this outburst the source was detected by Ginga in a low state,
making it the first NS binary in which a low soft X-ray state was 
definitively linked to a bright phase in hard X-rays. Our observations of 
4U~0614+091 extend these results, showing that the hard and soft fluxes are 
anticorrelated as the source varies. The spectrum of 4U~1608-522 exhibits a 
clear break at approximately 65 keV with spectral indices 1.8 and $3.2\pm0.2$.
For 4U~0614+091 the spectra may roll over at a lower energy, which would
have important implications for the hard X-ray emission models. This could 
point, for example, to lower electron temperatures in thermal models.

The present data indicate that when the RXTE pointed observations were 
performed, on TJD 10195 and 10197, the source was in a low state with a hard 
power law spectrum ($\alpha\sim2.1$) and little contribution from a soft 
component. During these observations we have discovered two high frequency QPOs
over the range 580 to 950 Hz (Ford et al. 1996c). The QPO frequency and total 
count rate are linearly correlated, while the frequency difference between
the two peaks is constant. Perhaps the presence of kHz QPOs is linked to the
hard X-ray brightness. The two other atoll sources with kHz QPOs,
4U~1728-34 and 4U~1608-52 (Strohmayer et al. 1996, van Paradijs et al. 1996),
have also been detected as bright hard X-ray sources by BATSE (Barret et al. 
1996).

\section{Conclusion}

Combining the capabilities of BATSE earth occultation monitoring and the
ASM on RXTE, we have measured the flux of 4U~0614+091 over a wide energy band 
(1--100 keV) during several months. We detected an active hard X-ray phase
accompanied by soft X-ray variability. The hard X-ray and soft X-ray
fluxes show a distinct anticorrelation. This measurement is an example of the 
new opportunities now available for broad-band X-ray monitoring.

\acknowledgments

We would like to acknowledge the BATSE instrument team and the members of the 
RXTE Guest Observer Facility for their support and assistance.
This work is supported in part by NASA Grants NAG~5-2235 and NGT~8-52806.

%%%%%%%%  TABLE 1: SPECTRAL FITS  %%%%%%%%%%
\begin{deluxetable}{llll}
\tablewidth{33pc}
\tablecaption{Spectral Fits}
\tablehead{
\colhead{Model: Data} & \colhead{$\alpha$} & $A_{0}$ [$\times10^{-3}$] &
 \colhead{$\chi^2$, $\nu$}}
\startdata
PL: BATSE           & 2.68 (0.35) & 7.05 (3.57) & 1.1, 3 \nl
PL: ASM             & 2.23 (0.33) & 2.30 (1.06) & 0.5, 1 \nl
PL: BATSE+ASM       & 2.09 (0.08) & 2.92 (0.32) & 1.1, 6 \nl
CUTOFFPL: BATSE+ASM & 1.79 (0.18) & 3.83 (0.97) & 1.8, 5 \nl
                    & $E_{c} = 45(20)$ keV & \nl
\tablecomments{Fits to data in the interval TJD 10178--10190.
The spectral fits are of the form, PL: $A_{0}(E/\rm{10~keV})^{-\alpha}$,
CUTOFFPL: $A_{0}(E/\rm{10~keV})^{-\alpha}e^{-E/E_{c}}$ 
(ph~cm$^{-2}$ s$^{-1}$ keV$^{-1}$). Errors (parenthesis) are $1\sigma$.}
\enddata
\label{table:spec_fits}
\end{deluxetable}

%FIGURES%%%%%%%%%%%%%%%%%%%%%
%%%% the following are the same figures scaled for the preprint size (aas2pp4)

\clearpage

\end{document}